\documentclass[journal]{IEEEtran} 
\usepackage{CJK}
\usepackage{amssymb}
\usepackage{amsmath}
\usepackage{graphicx, subfigure}
\usepackage{url, cite}
\usepackage{verbatim}
\usepackage{booktabs}
\usepackage{multirow}
\usepackage{bigstrut}
\usepackage{booktabs}
\usepackage{amsmath,cite,graphicx,times,subfigure,url,verbatim}
\usepackage{algorithmic}
\usepackage{algorithm}
\usepackage{epstopdf}

\usepackage{subfigure}
\usepackage{color}
\usepackage{cite}
\usepackage{enumerate}
\usepackage{array}

\newtheorem{subsec:coding}{subsec:coding}

\usepackage{color}

\begin{document}

\title{Ready Player One: UAV Clustering based Multi-Task Offloading for Vehicular VR/AR Gaming}

\author{
Long~Hu,~Yuanwen~Tian,~Jun~Yang,~Tarik~Taleb,~Lin~Xiang,~Yixue~Hao
\thanks{L. Hu, Y. Tian, J. Yang and Y. Hao are with School of Computer Science and Technology, Huazhong University of Science and Technology, China. (Email: hulong@hust.edu.cn, yuanwentian@hust.edu.cn, junyang\_cs@hust.edu.cn)}
\thanks{T. Taleb is with Aalto University, Finland, Centre for Wireless Communications (CWC), University of Oulu, Finland, and The Computer and Information Security Department, Sejong University, South Korea. (Email: tarik.taleb@aalto.fi)}
\thanks{L. Xiang is with University of Luxembourg, Luxembourg. (Email: lin.xiang@uni.lu)}
\thanks{Yixue Hao is the corresponding author. (Email: yixuehao@hust.edu.cn) }
}

\maketitle

\begin{abstract}
With rapid development of unmanned aerial vehicle (UAV) technology,
application of the UAVs for task offloading has received increasing
interest in the academia. However, real-time interaction between one
UAV and the mobile edge computing (MEC) node is required for processing
the tasks of mobile end users, which significantly increases the system
overhead and is unable to meet the demands of large-scale artificial
intelligence (AI) based applications. To tackle this problem, in this
article, we propose a new architecture for UAV clustering to enable
efficient multi-modal multi-task task offloading. By the proposed
architecture, the  computing, caching and communication resources
are collaboratively optimized using AI based decision-making. This
not only increases the efficiency of UAV clusters, but also provides
insight into the fusion of computation and communication.
\end{abstract}


\markboth{Under review: IEEE Network, VOL. XX, NO. YY, MONTH 20XX}{}

\section{Introduction}\label{sec:introduction}

\begin{figure*}
\centering
\includegraphics[width=6in]{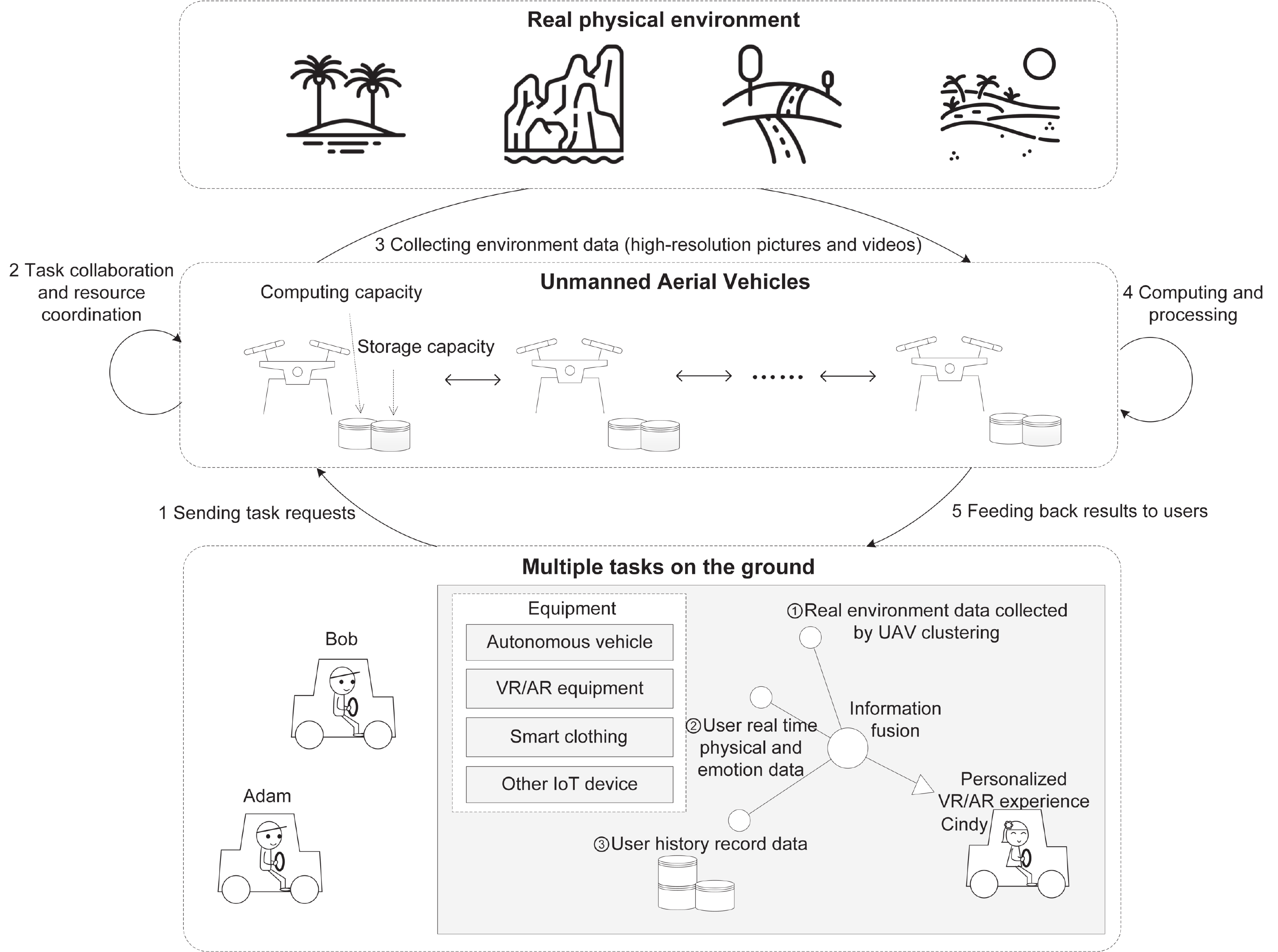}
\caption{Illustration for vehicular VR/AR gaming scene enabled by UAV clustering based multi-task offloading}
\label{fig1}
\end{figure*}

With rapid advances in mobile computing and wireless communication technologies, the demands of mobile end users have been largely met by deploying edge computing~\cite{Chen2018Task} and other solutions on the ground, while employing Internet of Vehicles as a supplementary infrastructure to accommodate unmanned applications~\cite{Chen2018Cognitive},~\cite{Taleb2018}. Recently, researchers have shown increasing interest towards processing in the air complex tasks as automatic cruise, aerial photography, and precision target identification. However, traditional architecture for enabling collaboration between one unmanned aerial vehicle (UAV) and mobile edge computing (MEC) is not applicable. Firstly, the miniaturization of UAV severely limits its computation, caching and communication (3C) capabilities. That is, for the same cost, the computing capabilities of UAVs are inferior to those of autonomous vehicles on the ground. Secondly, due to frequent interactions between the UAV and the MEC during processing tasks of mobile end users, the battery power of the UAV will be drained rapidly, resulting in low processing efficiency. Thus, the architectures of the ground-based systems cannot be directly applied in aerial systems, highlighting the demand for new computing architectures in UAV scenes.

Due to their flexible deployment, UAVs have attracted extensive research activities. For example, researchers have investigated cooperating UAVs for providing communication coverage. In particular, Motlagh et al.~\cite{Motlagh2018} investigate a UAV-aided MEC system for crowd surveillance scene. Mozaffari et al.~\cite{Mozaffari2016Efficient} propose an efficient deployment scheme for providing coverage to ground users by exploiting multiple UAVs as wireless base stations. Lyn et al.~\cite{Lyn2018UAV-aided} propose a UAV-aided hybrid network architecture to assist ground base station (GBS), which can exploit UAV-aided offloading for both throughput gains and cost savings. However, these works~\cite{Motlagh2018}\textendash\cite{Lyn2018UAV-aided} have not explored the role of UAV clusters nor the application of artificial intelligence (AI) technology. Other researchers have considered the problem of one UAV processing AI tasks such as disaster relief task, precision target identification, etc. For example, Zhao et al.~\cite{Zhao2018Saliency} propose a deep learning algorithm for applying a UAV to identify wildfire. However, for wide-range mountain fires in reality, e.g., the large-scale California mountain fires in November 2018, UAV clusters need to complete the disaster relief task in a timely manner. Schwarzrock et al.~\cite{Schwarzrock2018Solving} propose an efficient task allocation scheme for UAV clusters based on the swarm intelligence. The task investigated in \cite{Schwarzrock2018Solving} can be decomposed into computation, caching and communication to achieve the collaborative optimization of resources. The aforementioned works \cite{Motlagh2018}\textendash\cite{Schwarzrock2018Solving} have promoted the development of UAV technology. However, in the scenes of large-scale mobile users, heavy task load will lead to high delay. To tackle this challenge, we propose a UAV collaboration framework to offload multiple complex tasks and consider the coordination of computation, caching and communication resources~\cite{Chen2018Edge-CoCaCo}, where  the efficiency of UAV teams is maximized using AI based decisions.

We consider the  virtual reality/augmented reality (VR/AR) gaming scene as shown in Fig. 1. With the development of AI technology, the number of mobile users and the demand for high-quality user experiences are increasing rapidly. In the VR/AR hybrid gaming scene described in ``Ready Player One"~\cite{Cline2018Ready}, the driver and the passengers may enjoy a real-time experience from augmented visual effects while the car is moving at high speed in the physical environment. Wearable devices~\cite{Chen2018Wearable} can be utilized to augment user experience. However, as the user distribution is changing dynamically in real-time, deploying static/fixed edge computing nodes in the state-of-the-art networks fails to meet their computing demands. As a result, a large number of computing tasks may pile up in hot spots. UAVs flexibly deployed for tracing the mobile users provide a promising solution
to tackle this issue. Based on the real-time high resolution videos in peripheral physical scenes, UAVs can facilitate virtual scene processing and provide the users with personalized experience. However, UAVs are expensive and a large number of UAVs may cause strong mutual interference in the air. Therefore,  enhancing the efficiency of UAVs is crucial for meeting the requirements of large-scale mobile users and, at the same time, guaranteeing high-quality user experience. Several important characteristics of the proposed architecture are listed as follows.
\begin{itemize}
\item \textbf{Multi-task offloading:} The traditional scheme~\cite{Zhao2018Saliency} only considers a single UAV for processing a single task. For VR/AR applications, although the UAV-aided MEC system~\cite{Motlagh2018} can mitigate this problem, it depends heavily on the infrastructure and, hence, is not applicable herein. In contrast, by our proposed scheme, one UAV can serve multiple tasks. In particular, the results of each task can be partially reused to serve other tasks in an opportunistic manner. As a result, the proposed scheme can significantly enhance the efficiency of UAVs on a large scale.
\item \textbf{Collaboration of UAV clusters:} By our proposed scheme, one task can be jointly processed using multiple UAVs. The UAV network consists of multiple dynamic resources. Considering that each UAV may have different loads while processing different tasks, the computation resources of the idle UAVs  can be shared with to overloaded UAVs to improve resource utilization.
\item \textbf{Joint optimization of computing, caching and communication resources:} The completion of one task is successful only if sufficient computing, communication and storage resources are available.
The VR/AR task may easily fail in the traditional single-UAV scene as the  resources are fixed and limited. By considering UAV collaborations in multi-task offloading scenes,  the UAV clusters form a dynamic resource pool. Meanwhile, their computing, caching and communication resources can be shared with each other, in a dynamic and flexible manner, to balance the utilization of resource.
\item \textbf{AI based decision-making:} By the traditional scheme, the interaction between one UAV and MEC causes heavy overhead. When considering multiple-UAVs collaborative operation, each UAV must perceive and forecast the mobility of neighboring UAVs and the dynamic resources of the UAV network, before the task offloading decisions are made on this basis. This causes many open problems such as high delay, task failure. In fact, considering UAV cooperation in multi-task offloading scenes can achieve joint optimization of computing, caching and communication resources of UAV clusters. Moreover, AI based decision-making is crucial to enhance the utilization of available resources for maximization of the system performance.
\end{itemize}
 The contributions of this work are as follows.
\begin{itemize}
\item We investigate the fusion of computation and communication in UAV networks. The current research of mobile UAV networks has only focused on the communication aspects of UAVs or the task-processing capability of one UAV. Different from the UAV literature, we consider the entire large-scale applications of vehicular VR/AR gaming, and propose a new research direction by the fusion of two fields.

\item Moreover, we construct a novel architecture, called UAV-M3T, for UAV clusters to collaboratively perform different tasks. Under this architecture, the trajectory, task offloading and network resource allocation for the cooperating
UAVs within the clusters can be jointly optimized.

\item Finally, we propose an AI based decision-making framework to facilitate UAV cooperation and joint optimization of computing, caching and communication resources. In this framework, deployments of UAV clusters both in advance based on historical data mining and in real-time based on real-time perception are considered. Experimental evaluation reveals that our proposed strategy can effectively improve the collaboration of UAV clusters.

\end{itemize}
In the remainder of this article, we present the proposed UAV collaboration architecture for multiple task scenarios in Section~\ref{sec:architecture}. Moreover, in Section~\ref{sec:3Ccoordination} we introduce the resource coordination method for cooperating UAVs  and discuss its advantages. Furthermore,  the dynamic deployment scheme and its experimental evaluation are elaborated in Section~\ref{sec:strategy}. Finally, Section~\ref{sec:conclusion} concludes the paper and discusses some interesting future work.

\begin{figure*}
\centering
\includegraphics[scale=0.53]{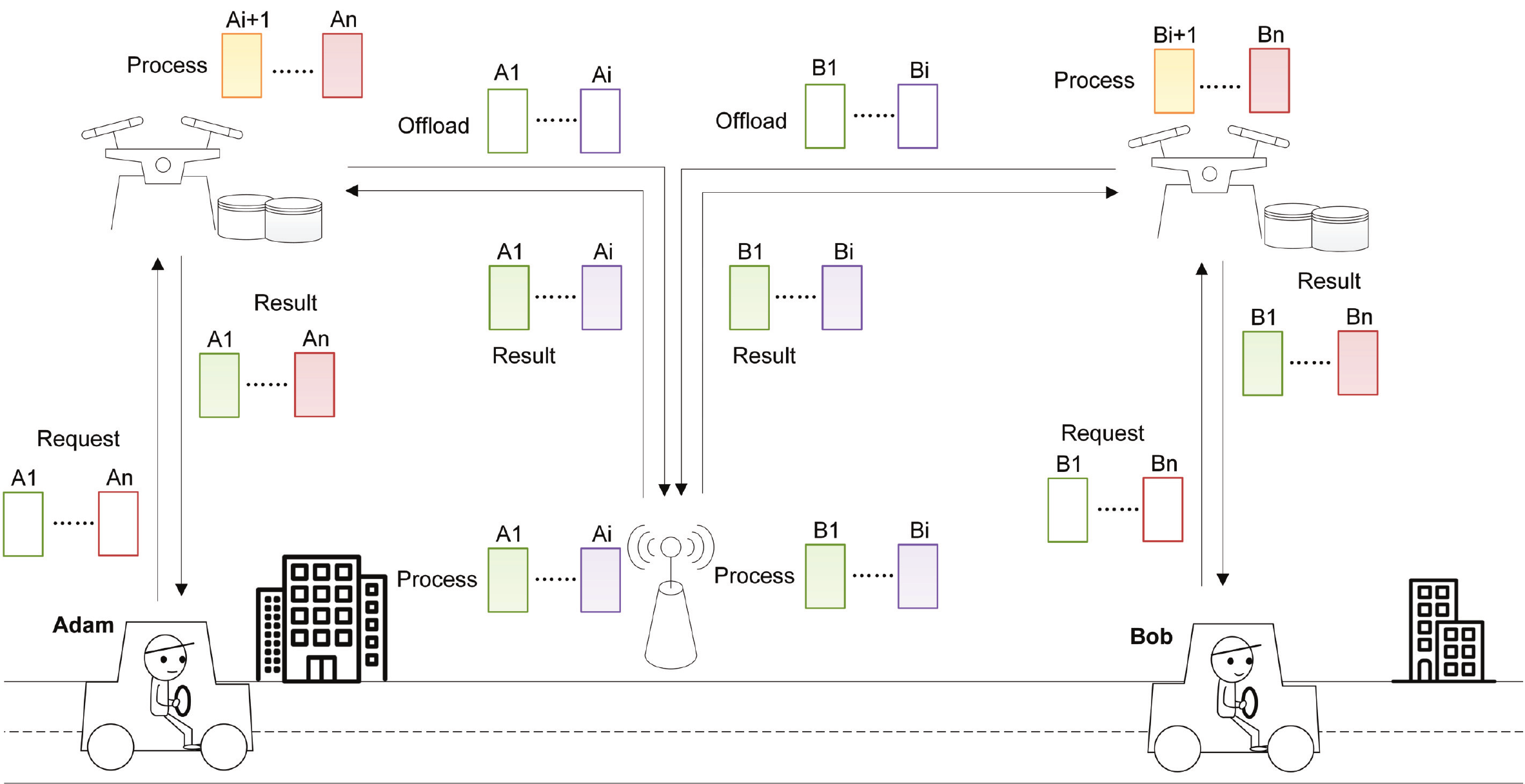}
\caption{Architecture of UAV-aided MEC task offloading}
\label{fig2}
\end{figure*}

\begin{figure*}
\centering
\includegraphics[scale=0.52]{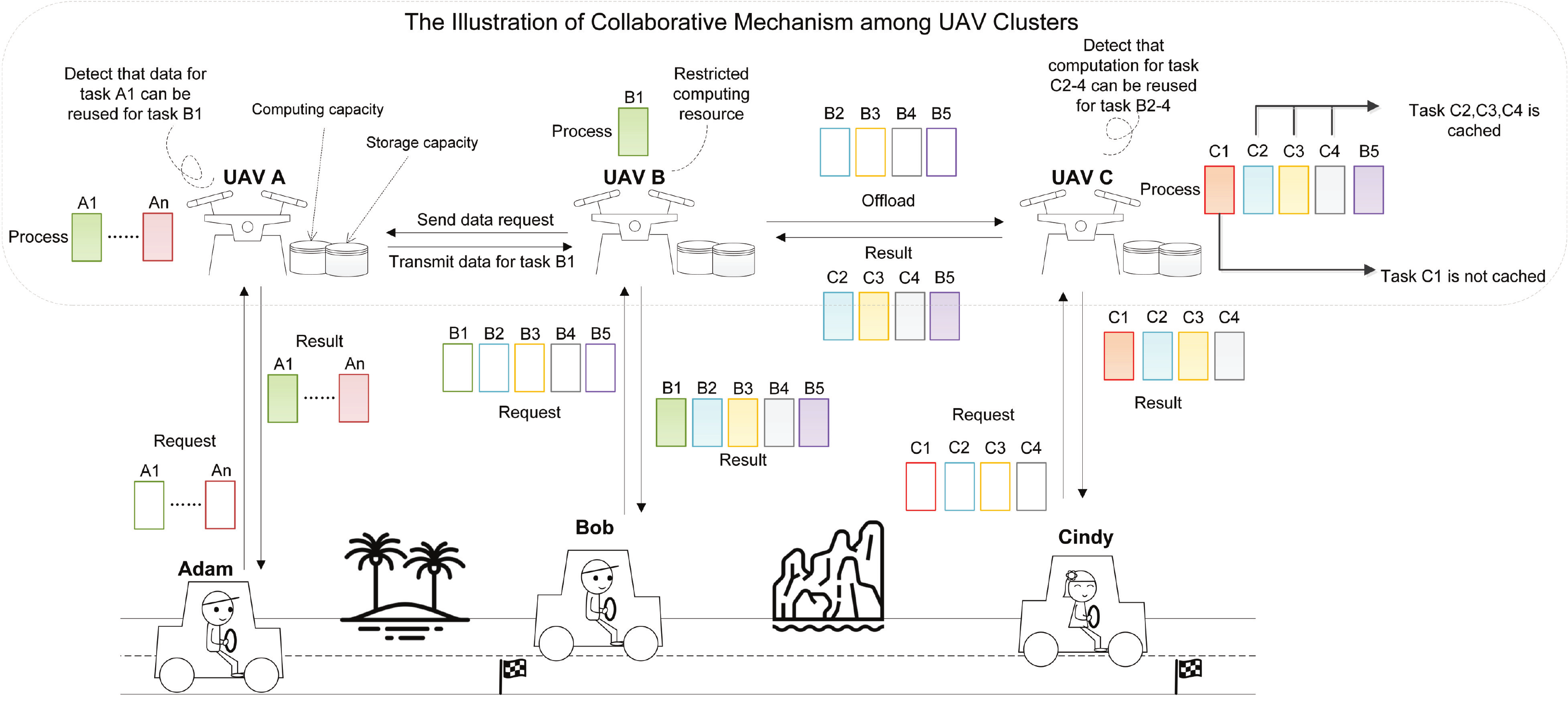}
\caption{Architecture of UAV clustering based multi-modal multi-task offloading (UAV-M3T)}
\label{fig3}
\end{figure*}

\begin{table*}
\begin{center}
\caption{Performance comparison of three architectures: MEC without UAV, UAV-aided MEC and UAV=M3T}
\begin{tabular}{|p{3cm}<{\centering}|p{2cm}<{\centering}|p{2cm}<{\centering}|p{2cm}<{\centering}|p{2cm}<{\centering}|p{1cm}<{\centering}|p{2cm}<{\centering}|}
	\hline
		Architecture&Deployment dynamicity&Resource flexibility&Real-time response&Intelligent decision-making&Cost&User's quality of experience\\
	\hline
		MEC without UAV&N/A&Medium&Medium&Limited&Low&Low\\
	\hline
		UAV-aided MEC&Limited&Medium&Medium&Limited&Medium&Medium\\
	\hline
		UAV-M3T&High&High&High&High&High&High\\
	\hline
       \end{tabular}
\label{tab1}
\end{center}
\end{table*}

\section{UAV clustering based Multi-Modal Multi-Task (UAV-M3T) Offloading Architecture}\label{sec:architecture}

\subsection{\label{sec2-1}Architecture of UAV-aided MEC task offloading}

Recently,  the task-processing mode of UAV-aided MEC networks
has been proposed in \cite{Motlagh2018}, whereby the computing task beyond a user's processing capability is offloaded to the UAV. The architecture is illustrated in Fig. 2. If the UAV has
limited computing capability available,  the task is then offloaded
to the ground MEC server for processing. In the latter case, the UAV
is used as a repeater to efficiently offload of the user's
computing task to the MEC server when e.g. the user has a poor communication
connection to the MEC server.

However,  the UAV-aided MEC networks may fail to meet the users'
required quality of experience in several key scenes. For example, for the VR/AR application scene in wild, desert, and complex topographies, the ground MEC network may not be conveniently and reliably built. If the ground MEC system is absent, the UAV-assisted MEC network architecture fails to promptly address
the situation. On the other hand, even if the ground MEC exists,
the users may distribute in a large area such that  it is difficult
to fully offload the tasks to the static/fixed MEC, which degrades the users' quality of experience. To tackle these issues, it is necessary to adopt a flexible task processing architecture based on e.g. UAV
clustering.

\subsection{Architecture of UAV-M3T task offloading}
We construct a novel architecture, called UAV-M3T, for UAV clusters to collaboratively perform different tasks. The architecture is illustrated in Fig. 3.
\begin{enumerate}
\item \textbf{UAV-O2O Mode (One UAV to One Task)}: The simplest mode in
UAV-M3T offloads one user task to one UAV that has sufficient computing,
caching and communication capabilities for processing. This mode
has the lowest  cost but can still fully exploit the advantages of
UAV clusters. We note that the one UAV-aided MEC service mode discussed
in Section \ref{sec2-1} is essentially a result of introducing the MEC server as backup
resources into the UAV-O2O mode.
\item \textbf{UAV-O2M Mode (One UAV to Multi-Task)}: The UAV-O2M mode differs
from the UAV-O2O mode in that the former does not process the tasks
separately, but can reuse the tasks fo improve the users' quality
of experience. An example of the UAV-O2M mode is illustrated in Fig. 3. If the UAV-O2O
mode is adopted, the tasks of users Adam, Bob and Cindy will be processed
by UAVs A, B and C, respectively. This significantly reduces the processing
efficiency of the UAVs. For example, the data collection tasks from a group of neighboring users within the same time window are usually the same. Therefore, the multi-user data collection task can be delegated to one UAV for saving computing resources. As shown in Fig. 3, since
Adam and Bob are in the same area, the data collected at UAV A can
be transmitted to UAV B for computing, while the computation results
of UAV B can be directly fed back to and used at both Adam and Bob.
In this way, the UAV-O2M mode utilizes the technique of task resuing.
\item \textbf{UAV-M2O Mode (Multi-UAV to One Task)}: In the UAV-M2O mode,
multiple UAVs collaboratively process one task. As shown in Fig. 3,
the VR/AR gaming task of user Adam is allocated to UAVs A, B and C for joint processing. In particular, the landscape data collected by UAV A is first transmitted to UAV B for processing. If UAV B has
only limited computing resources and fails to serve all the task requests
of Adam, a portion of the tasks will be then offloaded to UAV C. Finally, UAV C will utilize its idle computing resources to process the task of Adam jointly with UAV B. As a result, the M2O mode can efficiently utilize the network resources of the UAV clusters by enabling cooperation
among neighboring UAVs. This significantly improves the users' quality
of experience and leads to efficient resource allocation.
\item \textbf{UAV-M2M Mode (Multi-UAV to Multi-Task)}: The UAV-M2M hybrid
service mode combines the UAV-O2M mode and the UAV-M2O mode. The hybrid
service mode is the most common mode of UAV clusters cooperation in
processing multi-task scenes.
Multi-agent system has been investigated in communication systems~\cite{Fortino2005Multi-agent},
along with agent-based implementation on smart objects in IoT systems~\cite{Motlagh2016}.
By adopting the UAV-M2M mode, the trajectory, task offloading
and network resource allocation for the cooperating
UAVs within the clusters can be jointly optimized.
\end{enumerate}
Research into UAV-M3T architecture is promising for future AI based applications.
Although the UAV-M3T architecture in the hybrid service mode has a relatively
high deployment cost, it can significantly improve users' quality
of experience  and provide brand new market returns for the service
provider. Table I presents a comparison of the three architectures.
In fact, several projects on facilitating the UAV-aided MEC applications
have been recently launched by  Google, Facebook, Amazon and Huawei.
It is expected that the deployment cost of UAV clusters will be continuously
reduced in the future. Moreover, the advent of advanced Beyond 5G
(B5G) technology will facilitate a widespread deployment of UAVs to
meet users' rising requirements on  quality of experience.


\section{Coordination of Computing, Caching and Communication Resources}

\label{sec:3Ccoordination} 
The key performance indices of UAVs include capacity, delay, energy,
reliability, and cost, etc. The quality of experience measures customer's
satisfaction level, which depends on the personal preference of the user, environment and service. During the task processing, the actual tasks themselves are multi-modal.
Due to their heterogeneity, different tasks demand for different
computation, caching, and communication resources. In the proposed system, the deployment of computation, caching, and communication resources using UAV clusters has advantages in the following aspects.
\begin{itemize}
\item \textbf{Amount of information collected:} Even if UAVs serve
different independent objects, the collected information can be highly
redundant due to the requirements of the same business such as the
VR/AR gaming scene. Therefore, the data validity can be enhanced by
means of data reusing, content caching and task migration etc.
\item \textbf{Real-time performance:} For UAV-aided MEC architecture, a large quantity of information
collected by the UAVs needs to be transmitted back to the ground without
compression. This causes communication disruptions and fails the tasks when the bandwidth
is insufficient. For the multi-UAV clustering based collaboration architectures, many tasks
are compressed and processed in real-time during the UAVs' flight before been offloaded, therefore can reduce the communication delay of data transmission.
\item \textbf{Decision capability:}
Due to its limited computing, caching and communication capabilities,
a single small UAV can only support limited network decision-making.
UAV-M3T architecture can realize decision based on
network resources and mobility, as stated in Section IV. Next, for
the performance of decisions, the tasks with very high requirements
on performance can be completed by ensemble learning. However,
one UAV deploying the ensemble learning fails to meet the user requirements
of real-time performance due to high computing cost.
\item \textbf{Efficiency:} For complex application scenes such as VR/AR
gaming scenes, enhancing the efficiency of UAV will reduce costs.
On the one hand, efficient data collection and task processing can
be achieved by task reusing, content caching and other strategies.
On the other hand, UAV clusters collaboration including multi-UAV data
collection, resource allocation coordination, and intelligent decision-making
can enhance the overall resource efficiency of UAV clusters.
\end{itemize}

By considering the cooperation between UAVs in multi-user scenes, we can
achieve efficient  sharing of computation, caching, and communication resources among the UAVs to increase the system throughput.  Meanwhile,
the data and signaling exchanges between cooperating
UAVs can be reformed using e.g. device-to-device (D2D) connections.
The price is an increased transmission delay as the resources need
to be offloaded to other terminals using the D2D communication between
UAVs. Thus, in case of multiple users, the trade-off between  the
cooperation gains and the resulting system overhead needs to be investigated.
For this purpose, we assume that UAV clusters within the same organization are connected
by D2D and that one user's task is completed by a designated UAV. For notational convenience, we assume that only one user requests a VR/AR gaming task. Let $r_{i,j}^{t}$ be the communication data rate between UAVs $i$ and $j$. Moreover, $\beta_{i,j}^{t}$ and $\kappa_{i,j}^{t}$ are the amounts of computation offloading and caching content conveyed from the UAV $i$ to UAV $j$, respectively. For the considered VR/AR gaming scene, we optimize the average delay of the UAVs subject to the energy capacity of each UAV. We denote the UAV serving the requesting terminal as ``master'' UAV of the task and the UAV connecting the master UAV to provide 3C resources as the ``slave'' UAV. The delay $D^{UE}$ of the requesting user terminal accounts for both the average computating delay, which includes the computation delays in the master and slave the UAVs and the latency of D2D connection setup, and the average communication delay, which is the time needed to offload the tasks between different nodes. Moreover, $E^{UAV}_{i}$ denotes energy consumption of task computing and D2D transmission of $UAV_{i}$. $E^{MAX} _{i}$ denotes the maximum energy for $UAV_{i}$. The resulting resource allocation for UAV collaboration optimization problem is formulated as follows,

\begin{align*}
& \underset{r,\beta,\kappa}\min\quad D^{UE} \\
& \begin{array}{r@{\quad}r@{}l@{\quad}l}
s.t.& E^{UAV}_{i}&\leq E^{MAX}_{i}, &i=1,2,3\ldots,n\\
\end{array} .
\end{align*}

The solution of such optimization problem has been investigated in~\cite{Chen2018Label-less}. We can adopt online algorithms to determine
the optimal resource allocation when collaboration between UAVs is enabled. Furthermore, this problem formulating can be extended to include collaboration between UAV
clusters. When multiple tasks arrive simultaneously, the UAV clusters
can collaboratively optimize their 3C resources in the
same manner as above.


\section{Dynamic Deployment Strategy of UAV Clustering Based on Intelligent
Decisions}

\label{sec:strategy} 

\begin{figure}
\centering \includegraphics[width=3.5in]{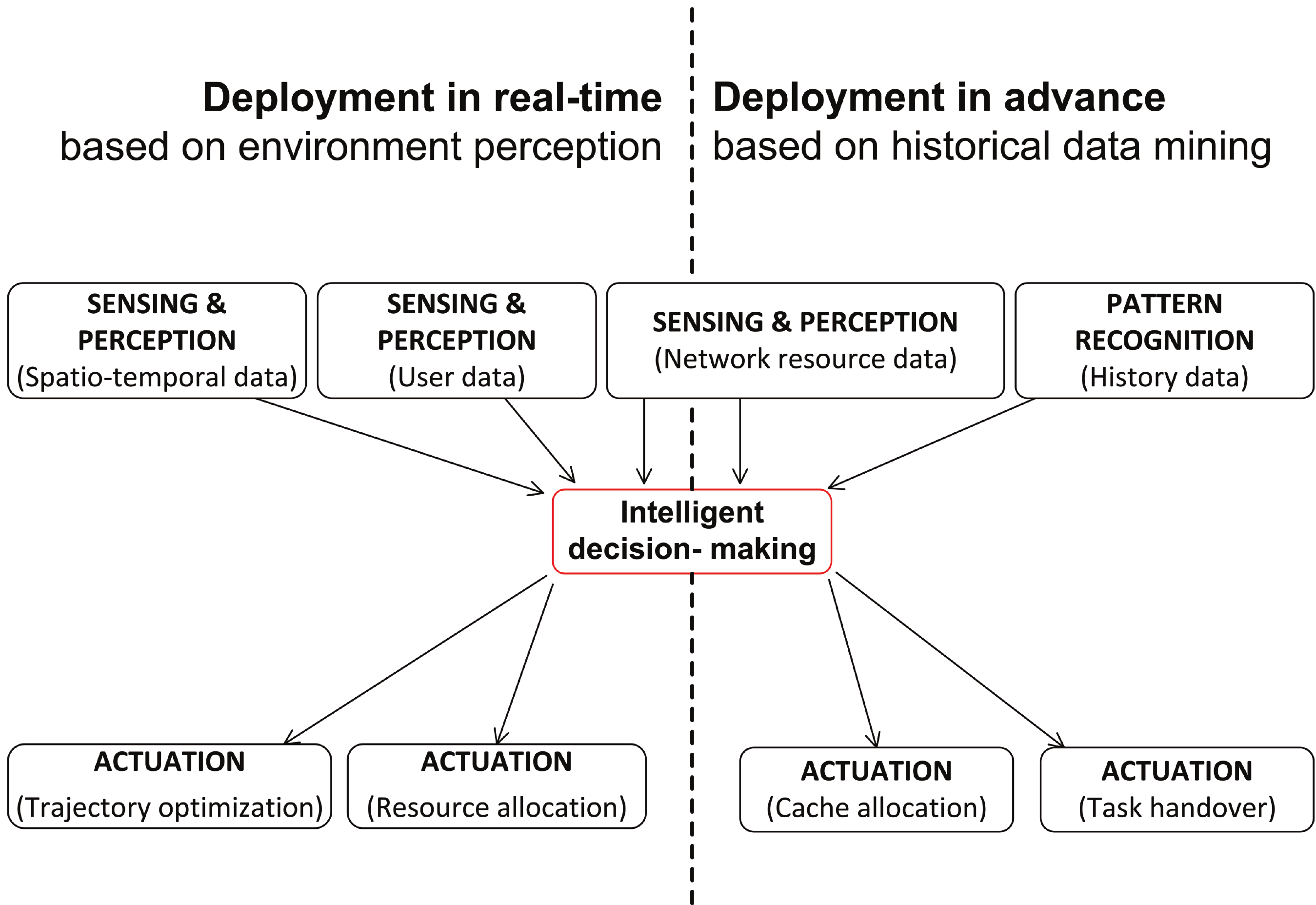}
\caption{Dynamic deployment strategy for UAV clustering}
\label{fig4}
\end{figure}

\subsection{Deployment in advance based on historical data mining}

The task assignment and resource allocation for UAVs can be deployed \emph{a priori} before the actual task requests are known, as shown in the right part of Fig. 4. This pre-deployment enhances the user experience because it can improve the network capacity and reduce the likelihood of
network congestion. For the VR/AR gaming scene, the tasks requested
by different users in the same location usually contain redundant information about the physical environment. Hence, historical and social data can be utilized via data mining to forecast the demands. If the data mining result indicates that a large number of users make similar task requests within a time window, the UAV can  cache the results and reuse them to serve the user demands at subsequent times. In this way, the delay and energy consumption are reduced simultaneously.

Next, the mobility of the UAVs can be forecasted periodically by collecting
the trace data of the UAV clusters in the historical time period to
optimize resource allocation. For 3C resource coordination in multi-task
scenes, the resources available at a given UAV and at its neighboring
UAVs should be jointly considered. When the mobility of UAVs is high,
the connection between UAVs may be interrupted due to increased likelihood
of link outage. In this case, dynamic adjustment of 3C resources is crucial
to improve the efficiency of resource allocation.

\subsection{Deployment in real-time based on real-time perception}

However, the historical and social data cannot accurately forecast
the user demands due to their dynamic nature. Thus, adaptive adjustment of the UAV clustering via real-time scheduling must also be made based on real-time perception to improve the network capacity and the user experience, as shown in the left part of Fig. 4. Sine each UAV in the UAV clusters may have a different path, the data perceived and the knowledge learned at the UAVs are different. In this way, the analysis of mutual information between UAVs can be enhanced using machine learning. When communication, computation and storage capacities change dynamically, the relevant real-time network status should be further analyzed in real time such that more UAVs will be sent
to the hot spots. The real-time collaboration and tracking optimization are conducted by several UAVs to balance the network resources. By considering multi-modal data, our optimization problem is generally non-convex due to non-convex constraints.
The traditional optimization scheme has long task duration. In view of high complexity of deep reinforcement learning, we should design the lightweight deep reinforcement learning algorithm for decision-making of UAV clusters. In our future work, we will try combing the Lyapunov optimization and deep reinforcement learning to further improve 3C resource allocation.

\subsection{Experiment: LSTM-based multi-UAVs load forecasting}

We adapt the real-time load forecasting and investigate the cooperation
and resource coordination among multiple UAVs. We then investigate on how multiple UAVs coordinate the resources with limited communication resource between each other. In real scenes, players need to be served by multi-UAVs
in UAV clusters simultaneously to carry out global resource scheduling on UAV clusters and estimate the load of the UAV nodes in the next time period. The UAV node defines a time series data, which can be forecasted using recurrent
neural network (RNN) model. It has been shown that RNNs are able to analyze deep semantic expression and time series information in data mining. However, RNNs suffer from a poor forecast capability when the resource load changes at large rate. To maintain the long-term memory of the RNN, we use
long short term memory (LSTM) network to eliminate the dependence of the forecast model on abnormal data. In the experiment, we forecast the changes of the UAV communication load state as an example.

\begin{figure}
\centering
\includegraphics[width=3.5in]{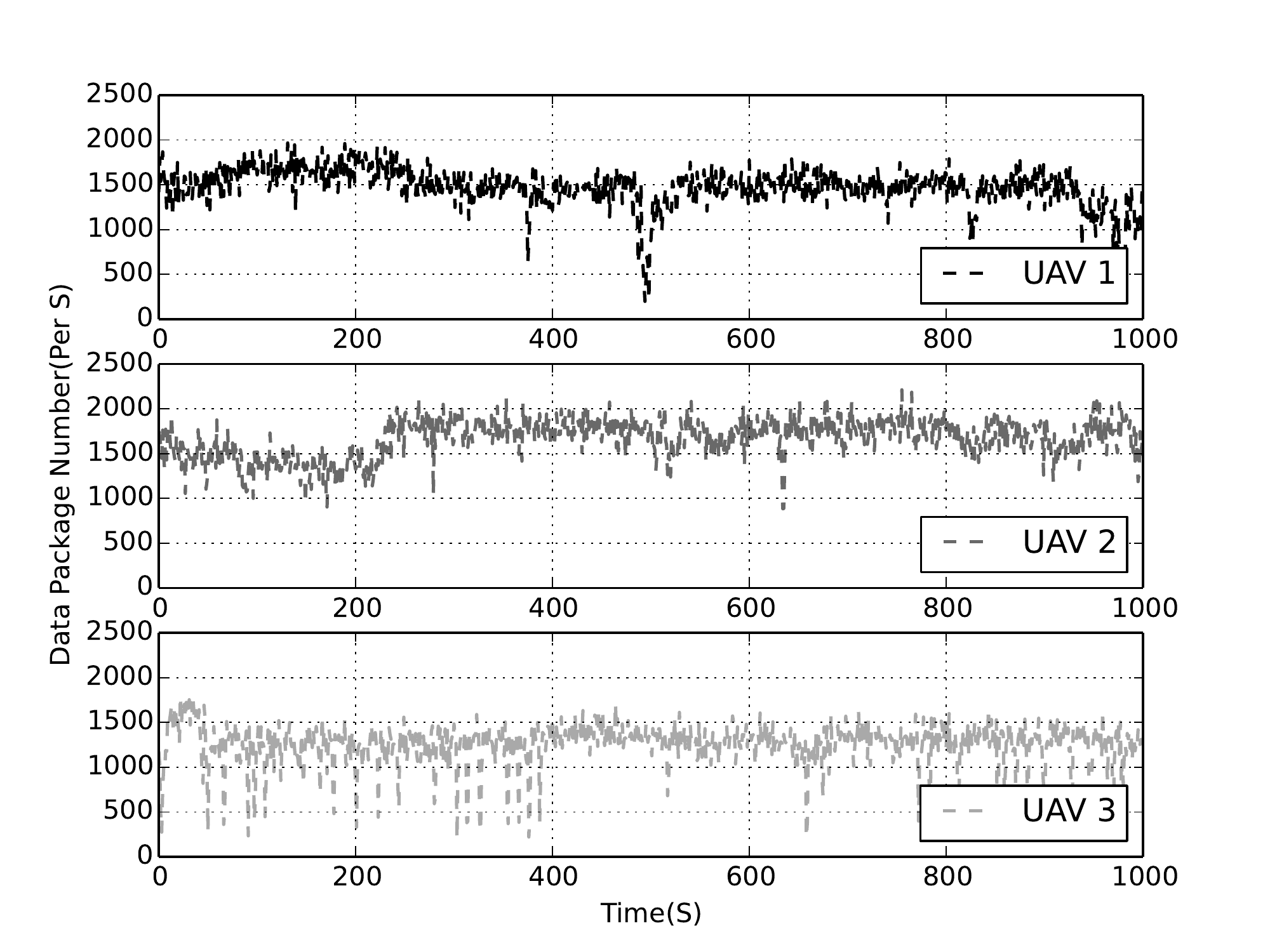}
\caption{Real-time task load forecasting for three UAVs collaboration}
\label{fig5}
\end{figure}

Fig. 5 shows the communication load state serving 3 UAVs to one player. The communication load changes of each UAV in the next time period are forecasted using the data of an hour prior to current time point as the reference, and forecast the current load change trend. After obtaining the communication load trend of each UAV, player divides the flexible task into a series of subtasks. Meanwhile, coordination of 3C resources is considered during task offloading to reduce the task delay and energy consumption. From our selected time window, around 250 second in particular, it can be observed that UAV 2 was trying to share some task load from UAV 1 while UAV 3 stayed relatively stable.

\section{Conclusion and future work}

\label{sec:conclusion} In this paper, we propose a new architecture, referred to as UAV-M3T, for vehicular VR/AR gaming. The UAV-M3T architecture utilizes AI based decision making for  collaborative optimization of the UAV team and the network resources and, hence, improves the task performance and resource efficiency of the UAVs. Our proposed scheme has extensive applications  in the military industry as well as city and business applications. However, many research challenges also need to be tackled. For example, we should consider improving resource coordination of UAVs in more complex scenes such as task migration~\cite{Chen2018Migration} and investigate efficient algorithms for dynamic deployment of UAV clusters, which are left as future work.

\section*{Acknowledgement}
This work was supported by the National Natural Science Foundation of China (Grant 61802138, Grant 61802139), the China Postdoctoral Science Foundation (No. 2018M632859).
This work was partially supported by the Academy of Finland 6Genesis Flagship (Grant No. 318927) and the Primo-5G project, that has received funding from the European Unions Horizon 2020 Research and Innovation Programme under Grant Agreement No.815191.
The work of L. Xiang is supported by the European Research Council (ERC) project AGNOSTIC.

\bibliographystyle{IEEEtran}

\end{document}